# FEATURE ENGINEERING FOR DATA-DRIVEN TRAFFIC STATE FORECAST IN URBAN ROAD NETWORKS.

**Felix Rempe (corresponding author)**
BMW Group,
Mobility- and Fleetintelligence, EE-541
Parkring 19, 85748 Garching, Germany
Tel: +49 151 601 96963, Email: Felix.Rempe@bmw.com

**Klaus Bogenberger**
University of the Federal Armed Forces Munich,
Department of Traffic Engineering
Werner-Heisenberg-Weg 39, 85579 Neubiberg, Germany
Tel: +49 89 6004 2503 Email: Klaus.bogenberger@unibw.de

Word count:  7,258 words text + 0 table x 250 words (each) = 7,258 words

Submission Date: 01.08.2018

**ABSTRACT**

Most traffic state forecast algorithms applied to urban road networks consider only the links in close proximity to the target location. However, for longer-term forecasts also the traffic state of more distant links or regions of the network are expected to be a valuable feature for data-driven traffic state predictors. This paper studies this expectation using a network clustering algorithm and one year of Floating Car (FCD) collected by a large fleet of vehicles. First, a clustering algorithm is applied to the data in order to extract congestion-prone regions in the Munich city network. The level of congestion inside these clusters is analyzed with the help of statistical tools. Clear spatio-temporal congestion patterns and correlations between the clustered regions are identified. These correlations are integrated into a KNN travel time prediction algorithm. In a comparison with other approaches this method achieves the best results. The statistical results and the performance of the KNN predictor indicate that the consideration of the network-wide traffic is a valuable feature for predictors and a promising way to develop more accurate predictors in the future.

*Keywords*: Traffic state prediction, Network Clustering, Floating Car Data, Congestion Patterns

## INTRODUCTION

Accurate traffic state forecasts in urban road networks constitute highly relevant information for different applications in traffic management, transport planning, etc. Though, prediction in urban networks face several challenges. Compared to freeways, an urban road network comprises thousands of segments which are connected via controlled or non-controlled intersections. Vehicles brake and accelerate frequently, depending on interactions with other vehicles, traffic signals or other traffic participants. Due to the network-structure, traffic congestion usually affects not just a linear road segment but may develop over large subnets of the original network. Additionally, the amount of available data in urban road networks is usually limited. Therefore, the development of accurate traffic state forecast algorithms is still a matter of current research.

The usual approach is to apply data-driven models, which render the explicit modeling of a system unnecessary. Many approaches tackle the challenge with variations of Auto-Regressive Moving Average (ARMA) method. To mention a few, (1) apply a multivariate spatial-temporal autoregressive model on a sample network with different road categories. (2) apply a STARMA model to estimate arterial traffic conditions, (3) model the traffic flow in space and time using a Space-Time Autoregressive Integrated Moving Average (STARIMA) model. (4) compute correlations between edges in the London traffic network in order to analyze required model complexities for models such as STARIMA. A broader review of currently applied methods is given in (5). Other machine learning methods apply Neural Networks (6, 7, 8), Support Vector Machines (9), Boltzmann machines (10) etc.

In (11) ten major challenges of traffic forecasting are pointed out which mark promising directions to increase the forecast accuracy. Among them is the need to consider temporal as well as spatial dependencies between traffic conditions on different edges in the network. Many approaches do so already, but only on a small scale. The forecast for one position in the network mostly depends only on adjacent edges of the network. For forecasts with a time horizon of minutes this is may be sufficient, though for longer horizons the traffic conditions throughout the entire network are expected to provide relevant information. One challenge is that a large region of influence means to process a large number of edges. In order to eliminate mutual dependencies, neighborhood selection techniques, such as a Graphical Lasso are proposed (12). (13) give a comparison about different methods. Still, these approaches are on an edge-level, plus, the neighborhoods are still local which does not allow to consider network-wide relations. In order to aggregate similarly behaving edges of a large network, (14) apply several clustering techniques. Though, edges do not need to be connected and, therefore, resulting clusters are disseminated. (15) propose algorithms in order to create so-called supernodes that represent connected subgraphs of large road networks. Edges of a subgraph are expected to have high similarities. The approach is validated with simulated data, but the number of resulting supernodes is still large. A different approach is to analyze congestion on a network-level. (16, 17) partition a road network dynamically into connected and congested subgraphs with similar properties. They observe the evolution of such congested regions over time and seek to determine a macroscopic fundamental diagram for urban networks.

The presented method in this paper is based on the work by (16) and extended with further concepts that allow for a more sophisticated congestion pattern analysis and forecast. One observation when applying the dynamic partitioning of the network is that congestion often emerges and resides in the same parts of the network. A possible explanation is that similar commuting patterns of travelers cause high traffic demands at the same bottlenecks every day, which ultimately leads to recurrent spatio-temporal congestion patterns. As a side-effect, there are also many edges of the network which are rarely or never congested. For traffic monitoring and prediction, these edges are less relevant and may be neglected in favor of a reduced model complexity and decreased computational times. The idea is to identify these regions in a network in which congestion occurs on a regular basis and focus monitoring and forecasting on said parts.

Besides a reduced computational effort, the identification of such regions has further benefits. One advantage is that the level of congestion at one bottleneck can be quantified with one variable. Compared to edge-based approaches where each edge has a different length and may be in a different traffic condition, an aggregation of all involved edges results in a more robust and representative variable for a subgraph. These aggregates computed for several congestion-prone parts of a network finally enable to analyze the spatio-temporal relations between traffic conditions at distant bottlenecks in the network. Additionally, a small number of variables is advantageous for analysis and visualization as well as for the training of data-driven forecasting methods with limited amount of data.

In the following chapters, first the available data and preprocessing is decribed. Next, the definition of the clustering algorithm is outlined. It is applied to the FCD and the resulting clusters are visualized. In a subsequent step, the traffic conditions inside these clusters are analyzed for congestion patterns. Finally, a data-driven traffic state prediction algorithm incorporating the inter-cluster correlations is proposed and compared to other methods.

## DATA PREPROCESSING

This chapter describes the available data and its preprocessing. A fleet of vehicles reports position data and according timestamps on a regular basis to a central server. Each vehicle that is equipped with a certain software version samples the current GNSS position in intervals of 10s to 30s. These positions and according timestamps are stored in the local memory of the vehicle. After sampling a few positions, a filter mechanism decides whether sampled GNSS positions are transmitted to the central server. This filter continuously compares the vehicle's velocity with the velocity given by a commercial traffic provider. If the velocity at one of the sampled data points deviates more than 10% to 30% (depending on the software version of the module) from the provided velocity, the recently sampled positions and according timestamps are transmitted to the central server. In case only small deviations are detected, the recently sampled positions are retained and removed from the local memory. Each transmitted position is linked to an alias that is generated by the vehicle. The alias is random and changes over time. Server-side, single transmitted positions of the same alias can be connected in order to reconstruct vehicle trajectories, which protects the driver's privacy.

Collected raw data is map-matched to a digital map. This map is represented as a directed graph:

$$G = (V, E) \tag{0}$$

which is an ordered pair of a (finite) set of vertices $V$ and a set of edges $E$. An edge $e \in E$ comprises a pair of two vertices $v_1, v_2 \in V$. In a directed graph, this pair is ordered such that there is a connection from vertex $v_1$ to $v_2$ but not necessarily from $v_2$ to $v_1$. The length and regulatory speed-limit of a road segment and the corresponding edge are denominated as $l_e(e) \in \mathbb{R}_+$ and $V_{Lim}(e) \in \mathbb{R}_+$, respectively. (More information regarding the map-matching process can be found in (18, 19, 20).

In total 318 days of data are available, resulting in approximately 400,000 velocity measurements $V_{Rec}(e, t)$ per day on a network comprising 17,413 major road edges (corresponds to 1826 km). For each velocity measurement, a relative speed-limit-based velocity is computed:

$$V_{Rel}(e, t) = \frac{V_{Rec}(e,t)}{V_{Lim}(e)} \tag{1}$$

In order to aggregate the measurements of individual vehicles into a space-time continuous traffic estimate and fill gaps for which no data is reported, the following considerations are done: First, time is discretized into intervals of $\Delta T = 1$ min, such that $T_d = 1, \dots, 1440$. Then, all velocity measurements $V_{Rel}(e, t)$ are interpreted as macroscopic traffic velocities for the respective time interval. If there are several measurements for the same time interval and the same edge $e$, their arithmetic average is computed. Since usually not for all time slices a velocity measurement is available, there are gaps in time and space. A simple estimation algorithm is applied: Each measurement is assumed to be valid for either 15min or until a new measurement is reported. This allows to have a mostly continuous representation of traffic conditions for peak hours of the day. For all remaining gaps, which appear mostly at night or on minor roads, free flow conditions estimated as $V_{Lim}(e)$ are assumed.

## CLUSTERING AND FEATURE ENGINEERING
This chapter describes the clustering of the network into connected subgraphs, which is the basis of the developed traffic prediction features. Note that the clustering is partly published in (21), which the interested reader can refer to for more details.

### Motivation
Feature extraction is an important step in the development of data-driven models. It helps to reduce the dimensionality of the input data to a smaller set of expressive variables which conclude the relevant information of the input data. This allows to filter unnecessary variables and get a better understanding, which information are relevant for the derived model. There are automated methods such as a Principal Component Analysis (PCA) which can be applied without domain knowledge. Though, the manual design of features, which is called features engineering, allows to integrate specific knowledge of the system into the features, which may not be visible in the noisy data of a limited set of data.

The properties of good features for a traffic state predictor are that the current value of a feature is related to the future state of the system. For instance, if one could measure the total number of vehicles with a certain travel time until they reach an infrastructural bottleneck, this feature would be very expressive to be evaluated whether the road will get congested or not. It would be especially useful, when a lot of data has already been collected which relate the number of vehicles and the state of congestion at the bottleneck.

The feature design in this case is based on the following consideration: It is assumed that recurrent congestion stems from recurrent commuting patterns of travelers who go to, or come from work. At times with low overall traffic demands, the infrastructure's capacity suffices to meet the demand. Traffic conditions are free in all parts of the network. With increasing demand more vehicles intend to pass certain bottlenecks of the network. When the bottleneck's capacity is exceeded, traffic gets congested and the Travel Time Loss (TTL) in these regions rises. With more vehicles entering the bottleneck region, more edges get congested and the TTL in said regions increases. Thus, the level of congestion of a congestion prone region relates to the current traffic demand on the corresponding bottleneck. Observing not only one but all bottlenecks in the network at the same time, and specifically their levels of congestion, this is expected to give a picture of the traffic demand of the entire network. Naturally, this is just an approximation of the real demand of the network. Since traffic demand is the physical cause for congestion, these levels of congestion in the bottleneck regions are expected to be promising features for traffic state forecast methods.

The approach is to identify these congestion-prone regions using historical FCD and aggregate the level of congestion in each of these regions into one variable. A clustering method initially described in (16) and extended in (21) is designed to extract these regions from data. The so-called clusters which result from the algorithm, have the the following properties:
1. They span the regions that are frequently congested
2. They are static over time
3. All edges of one congestion cluster are connected

In the following the cluster generation is summarized.

**Definition**
*Dynamic Clustering*

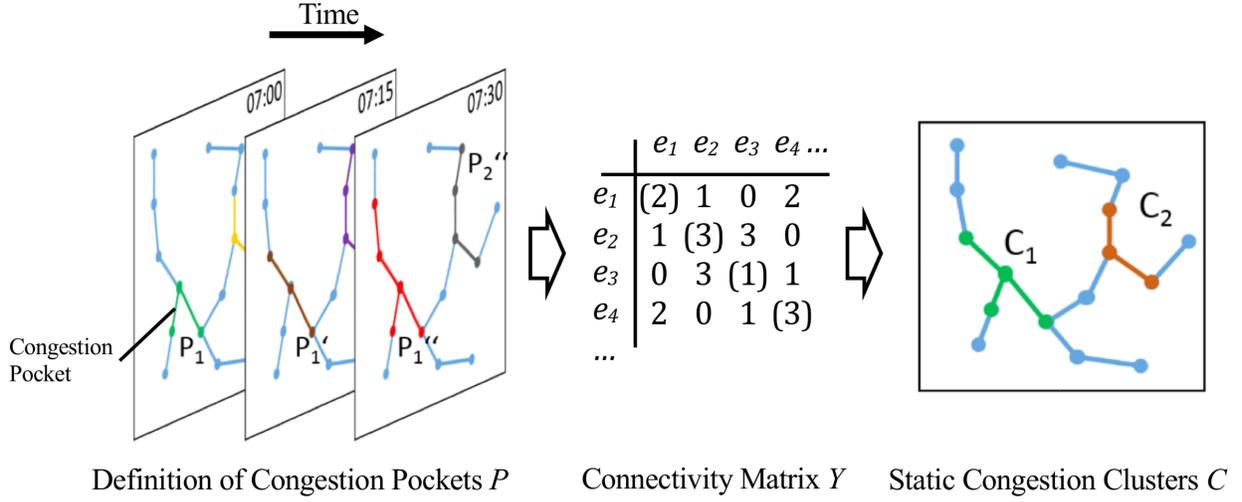

**FIGURE 1 Overview of the steps taken to process FCD into static congestion clusters (see (21))**

First, for each time step an edge is determined as free or congested depending on reported velocity data and the edge's speed limit. The function $J(e,t)$ ('J' for 'jam') is defined that indicates whether edge e is congested at time t:

$$J(e,t) = \begin{cases} 1 & if \; V_{Rel} \leq V_{Rel}^{thres} \\ 0 & otherwise \end{cases} \tag{2}$$

Similar to (16) a dynamic congestion pocket $P$ is defined. A congestion pocket denotes the spatial extent of an occurring traffic jam at a certain time $t$. Each congestion pocket is a time-dependent subgraph $G' = (V', E')$ of $G = (V, E)$ i.e. $V' \subseteq V, E' \subseteq E$. For each point in time the number and size of the congestion pockets may change. Formally, for some time $t$ and a congested edge $e^*$, a congestion pocket $P(e^*, t)$, is defined as the set of all edges $e \in E$, that have the subsequent properties:

1. $J(e,t) = 1 \; \forall e \in P(e^*, t)$
2. Either there exists a path from $e^*$ to $e$ on $G$, or a path from $e$ to $e^*$ on $G$ that consists solely of edges to which $J(e,t)$ assigns a value of one.

Thus, $P(e^*, t)$ describes a set of edges that are associated with edge $e^*$ which is part of the congestion pocket. For non-congested edges this set is consequently empty. This relation from edge to congestion pocket is surjective, which means that each edge at time t is mapped to one congestion pocket at maximum. On the other hand, a congestion pocket may refer to several edges.

*Static Clustering*
If two edges are frequently congested at the same time and they have a high proximity it is likely that those edges belong to a congestion-prone region at the same bottleneck. A static congestion cluster is supposed to agglomerate these edges. Dynamic congestion pockets model the proximity and congested state of edges such for one point in time. In the following, the temporal clustering of congestion pockets is described.

Assume that for all discrete time intervals $T = \{T_0, T_0 + \Delta T, \ldots, T_1\}$ congestion pockets are computed. The function $Y(e_1, e_2)$ is defined counting the number of time intervals in which two edges $e_1, e_2$ are part of the same congestion pocket:

$$Y: E \times E \to \{0, 1, \ldots, |T|\} \tag{3}$$
$$Y(e_1, e_2) \coloneqq |\{t \in T: e_1 \in P(e_2, t)\}| \tag{4}$$

The resulting quantities can be represented as a matrix (compare Figure 1). It is symmetric with the total number of time steps for which an edge is congested on its diagonal. For the clustering, the duplicate entries (due to the symmetry) and the diagonal elements are not relevant. Therefore, in the following matrix $Y^*$ as the strictly lower triangular matrix of $Y$ is considered. Two edges with a relatively high corresponding entry in $Y^*$ are called edges with high connectivity.

All edges with a high connectivity are supposed to be clustered into a finite number of static clusters $C_i \subseteq E, i = 1, \ldots, n_c$. An iterative algorithm is applied which assigns edges to different clusters $C_i$ based on $Y^*$. In short, the algorithm finds the pair of edges with the highest connectivity in the matrix. It checks, whether one of the edges is already assigned to any cluster. If not, a new cluster is defined that comprises these two edges. If one of the edges is already assigned to a cluster, the other edge is assigned to the same cluster. If both edges are already assigned to differing clusters both clusters are merged. Finally, the connectivity of these two edges is set to zero and the algorithm evaluates the next pair of edges. This procedure is done as long as:

$$\max\left(Y^*(e_1^*, e_2^*)\right) > Y_{min} \tag{5}$$

with

$$Y_{min} = \alpha \cdot \max(\{Y(e_1, e_2): e_1 \neq e_2\}) \tag{6}$$

Parameter $\alpha \in [0, 1]$ is introduced to decouple the clustering from the number of analyzed time intervals $T$.

In (21) different alpha parameters are compared and evaluated. Based upon this study, $\alpha = 0.15$ is chosen, which provides a decent compromise between the covered region of the network and the specificity of the clustered region. Due to the commuting behavior of the travelers, including the morning and afternoon rush-hour, it turned out that the congestion-prone regions are located in different parts of the network. Therefore, the cluster generation and analysis is divided into the morning period (00.00 – 12.00) and the evening period (12.00 – 24.00). This can be well explained with the position of infrastructural bottlenecks, which depend on the travel direction. Due to similarity of the results and shortage of space, in the following only the results for the morning period are presented.

Figure 2 depicts the ten largest clusters, which cover a length of 96.2km (52% of all clustered distance) during morning and 95.7km (55 % of all clustered distance) during evening hours.

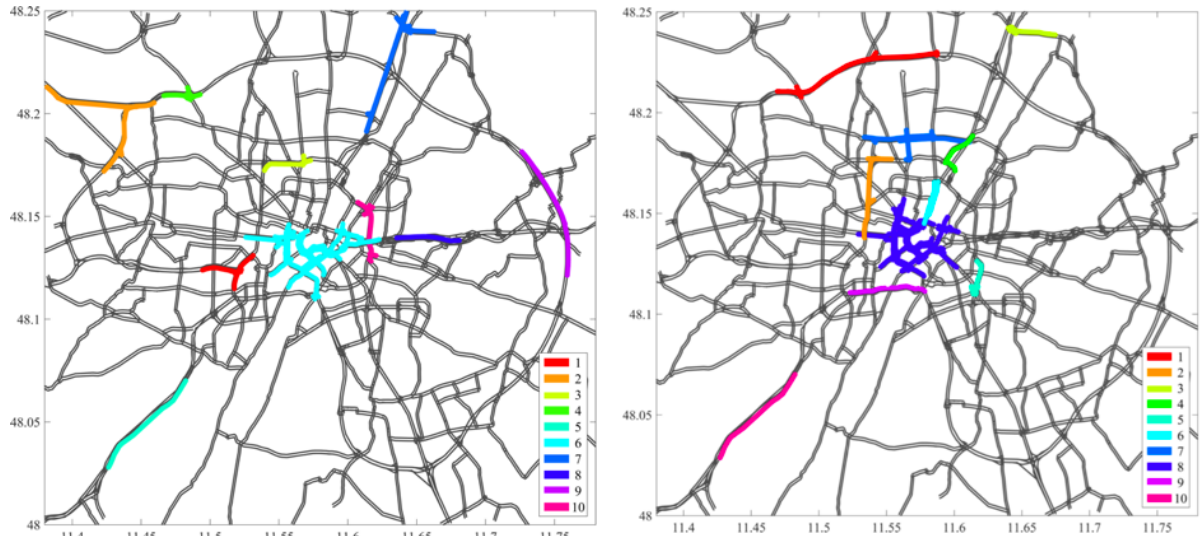

**FIGURE 2  Static clusters chosen for the following congestion pattern analyses (compare to (21)).**

## CONGESTION PATTERN ANALYSIS

In this chapter the traffic conditions inside the clusters are analyzed with statistical tools. The goal is to give an intuition why the observation of the bottleneck regions is valuable for traffic state forecasts.

### Level of Congestion

The level of congestion $\kappa(E^*, t) \in [0,1]$ describes the part of a cluster which is congested at time $t$:

$$\kappa(E^*, t) = \frac{\sum_{e \in E^*} J(e,t) l(e)}{\sum_{e \in E^*} l(e)}, E^* \subseteq E \tag{7}$$

Figure 3 visualizes exemplary the median $\kappa(E^*, t)$ of all clusters over all 318 days for the morning period.

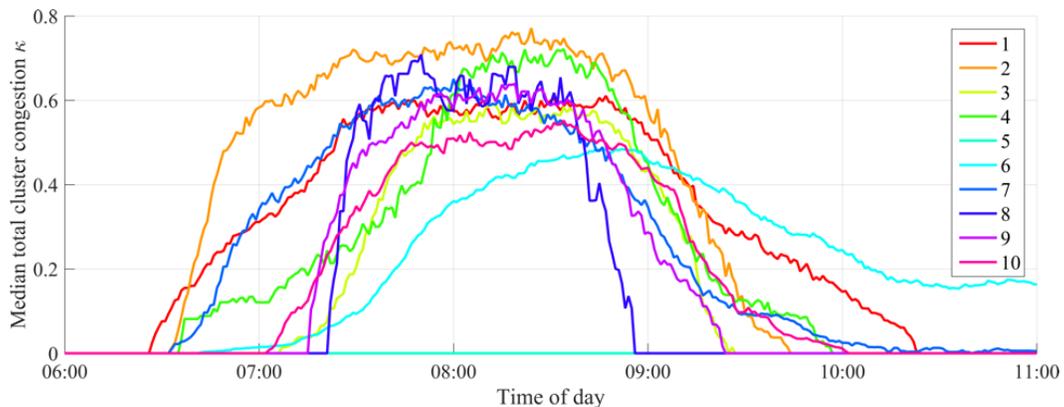

**FIGURE 3  Median cluster congestion of each cluster for morning and evening peak**
**           for all Mondays-Thursdays in 2015**

As visible 9 out of 10 clusters get congested during morning period in the median case. An outlier is cluster 5 which is located south-west of the city. An interesting fact is that not all clusters get congested at the same time (in average) but there seems to be a lag of up to one hour between the start of congestion in the morning.

**Cross-Correlation**

The consideration of the cross-correlations is a simple way to compare two signals that may have a temporal displacement and determine whether these signals follow a similar behavior. For congestion pattern analyses this allows to compare the congestion in different clusters including a temporal shift between starts and ends of congestion. For discrete signals the (normalized) cross-correlation is defined as:

$$R_{X,Y}(\eta) = \sum_{-\infty}^{\infty} Y[i]X[i+\eta] \tag{8}$$

$$\hat{R}_{X,Y}(\eta) = \frac{R_{X,Y}(\eta)}{|X||Y|} \tag{9}$$

where $|.|$ denominates the Euclidean norm. It can be applied as a flexible tool for data exploration that allows to obtain a first general overview of potential dependencies.

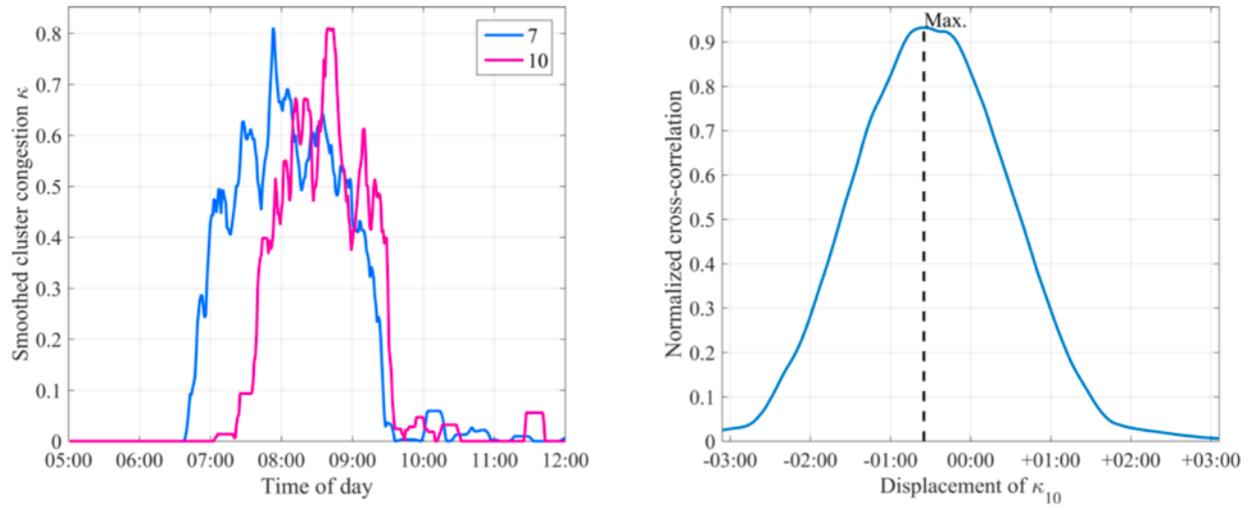

**FIGURE 4  Normalized cross-correlation coefficients between congestion in cluster 7 and cluster 10 on July 14th, 2015**

Figure 4 visualizes exemplary the level of congestion of two clusters (7 and 10) during morning hours and the resulting normalized cross-correlation for one day. It can be seen that the maximum cross-correlation for $\kappa_{10}$ is located at about -40minutes. This means that it its values of the time-series need to be shifted by 40minutes back in time in order to have the largest overlap with $\kappa_7$.

Formally, the lag between two signals is defined as:

$$\eta^{max} = \underset{\eta}{\operatorname{argmax}} \hat{R}_{X,Y}(\eta) \tag{10}$$

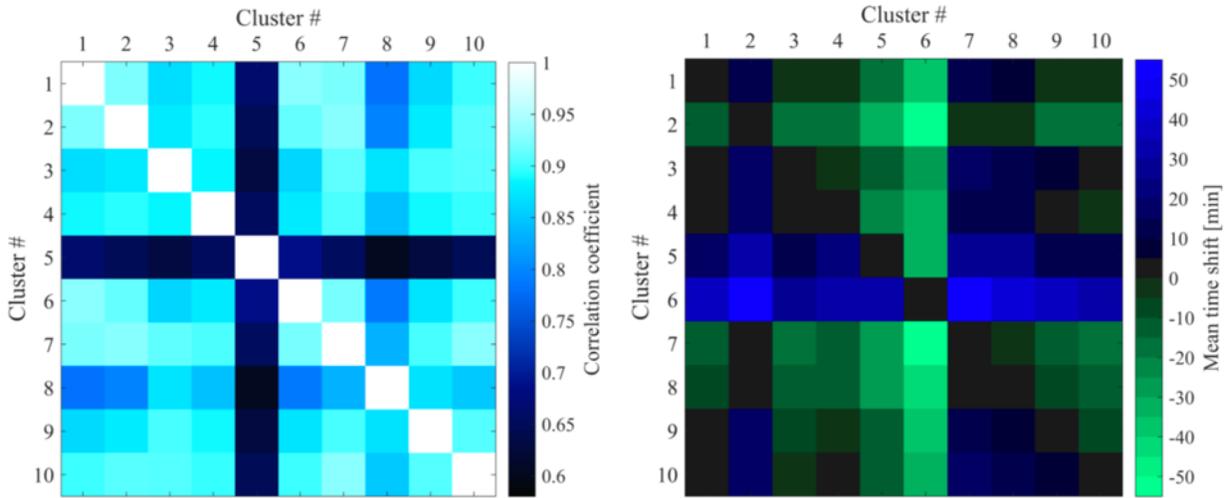

**FIGURE 5  Mean normalized cross-correlation and mean time shift between morning congestion inside congestion clusters over all workdays in 2015. (Notice: A positive lag for a cluster combination (i,j) means that congestion in cluster i starts later than congestion in cluster j)**

Figure 5 illustrates the cross-correlation coefficient and the average lag between all clusters averaged over 260 weekdays in 2015 during morning peak. Several observations can be made:

1. Having coefficients greater than 0.8 it can be stated that many clusters correlate strongly. Due to the overall peak hour and the median cluster congestions this result meets the expectations.
2. Cluster 5 correlates less with all other clusters. This matches with the previous analysis that cluster 5 behaves differently than the others.
3. Partially, there are strong lags between the congestion patterns: In average, congestion in cluster 6 is delayed about 40min to 50min compared to other cluster congestions. Clusters 2 and 7 tend to get congested earlier than other clusters. This also matches the previous findings.

To summarize, the cross-correlation analysis is a tool that provides an overview of the similarities between the level of congestion in the clusters. As such, it points out which clusters behave more similarly and which tend to get congested earlier/later than others. Though, its results may be misleading: If e.g. the duration of one congestion is significantly longer than the other one, the resulting cross-correlation is low since the maximal overlap between the two signals is reduced. Furthermore, the lag may be less obvious when the signals are complex. Therefore, cross-correlation coefficients and lags are only indicators of dependencies. Further analyses are required in order to verify hypotheses based on the cross-correlations.

**Daily Start and End of Congestion**

The previous studies revealed that, in average, there are correlations and lags between the time series of congestion among several clusters. In this section, the dependencies between congestion starts and ends on a daily basis are analyzed. The question that is investigated can be summarized as follows: 'Does the observation of congestion in one cluster allow to predict the start congestion in another cluster?'

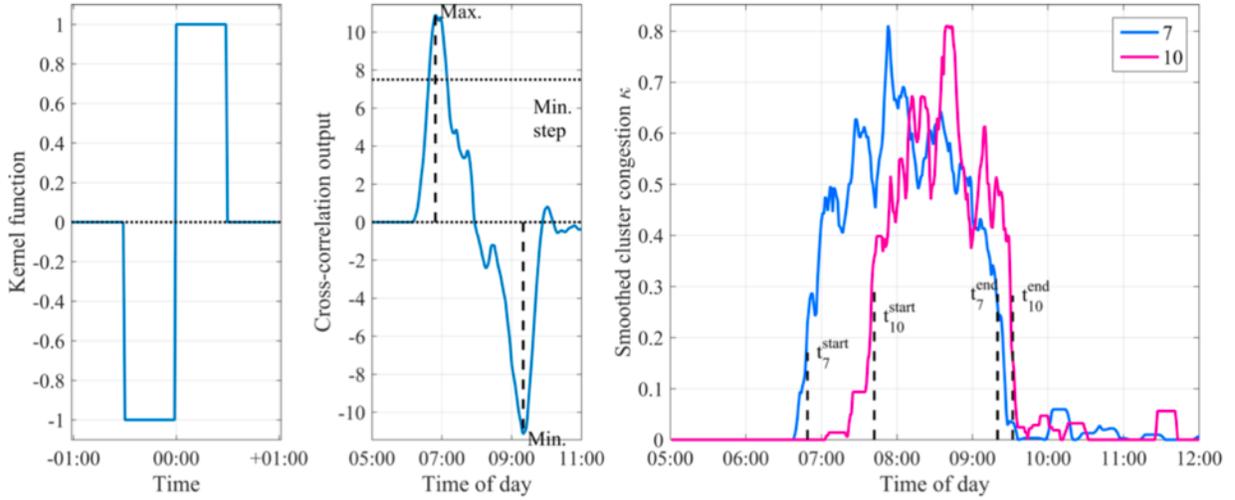

**FIGURE 6  Definition of congestion start and end times. Left: The applied kernel function, mid: the result of the cross-correlation with κ of cluster 10 observed on July 14th, 2015, right: the resulting start and end times of congestion for clusters 7 and 10**

In order to analyze starts and ends of congestion a definition of these events is required. Qualitatively, a congestion start is defined as the point in time when the level of congestion increases from a mean low level to a mean high level. Using the cross-correlation and a 1D-filter function for edge detection (see Figure 6), the start times of congestion are defined as follows:

$$t_i^{start} := \begin{cases} \underset{\eta}{\arg\max} \ R_{X,Y}(\eta) & \text{if } \max\left(R_{X,Y}(\eta)\right) > R^{thres} \\ 0 & \text{otherwise} \end{cases} \tag{11}$$

Where

$$S(t) = \begin{cases} -1 & if \ 0 > t > -T_A \\ 1 & if \ 0 < t < T_A \\ 0 & otherwise \end{cases} \tag{12}$$

And $T_A, R^{thres} \in \mathbb{R}_+$ denominate the time window used for averaging and a threshold that filters minor congestions, respectively. Congestion ends are defined similarly. Here the minimum cross-correlation valiue is considered and a previous congestion start must be existent in order to extract a congestion end.

Figure 6 illustrates the definition with $T_A = 30min$ min and $R^{thres} = 7.5$ and the resulting start and end times for congestion levels of cluster 7 and 10 on one exemplary day. Note that the definition utilizes the un-normalized cross-correlation. A threshold value of $R^{thres} = 7.5$ in combination with $T_A = 30min$ means that the average increase of congestion at the time of congestion start needs to exceed $7.5/30 = 0.25$ in order to be classified as a congestion start.

In the following the starts and ends of congestion of several pairs of clusters are discussed. Examples of clusters-pairs with significant lag in congestion, pairs with high correlation and pairs with low correlation are presented exemplary.

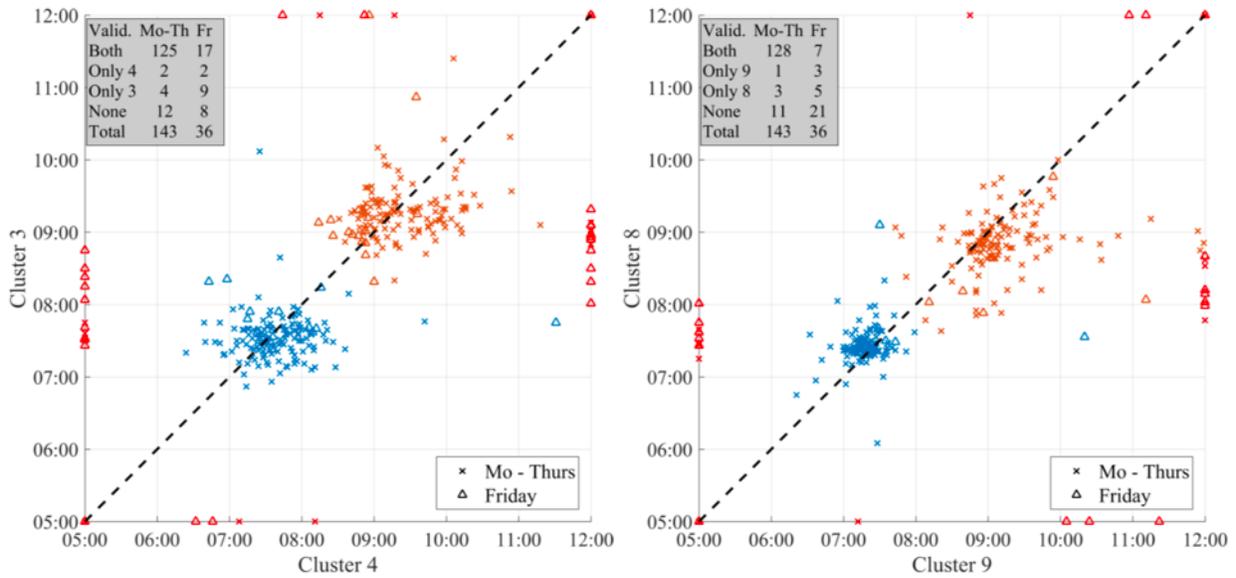

**FIGURE 7  Congestion starts and ends of several cluster combinations with high correlation and minor lag. Blue markers indicate the start times and orange markers the end times of congestion. Missing congestion are visualized as symbols on the axis of the diagram.**

Figure 7 illustrates two examples of cluster pairs with high correlation and minor lag for a total of 179 workdays in 2015 (due to technical reasons not for all working days data is available). For each day one (blue) marker indicates start time $t^{start}$ of congestion and one (orange) end time $t^{end}$. Two types of symbols distinguish between Monday- Thursday and Friday. If no congestion start and end for a cluster is identified for a certain day, these markers are drawn on the axis of the diagram. The dotted, isochronal line is a visual aid. A table in the upper left counts the numbers of starting/ending times with respect to different combinations of congestion status. Several interesting observations can be made:

1. In most days, both, cluster 3 and 4 get congested between 7am and 8am and congestion ends around 9am.
2. Outliers occur relatively often on Fridays; during Monday-Thursday outliers are relatively rare.
3. The distribution of congestions starts in cluster 8 and 9 is much narrower. It spreads around 7.30am.
4. Both cluster pairs are located in similar parts of the network, though not on the same road. The narrow distributions of starts and ends and the low number of outliers indicate a strong dependency between these bottlenecks.

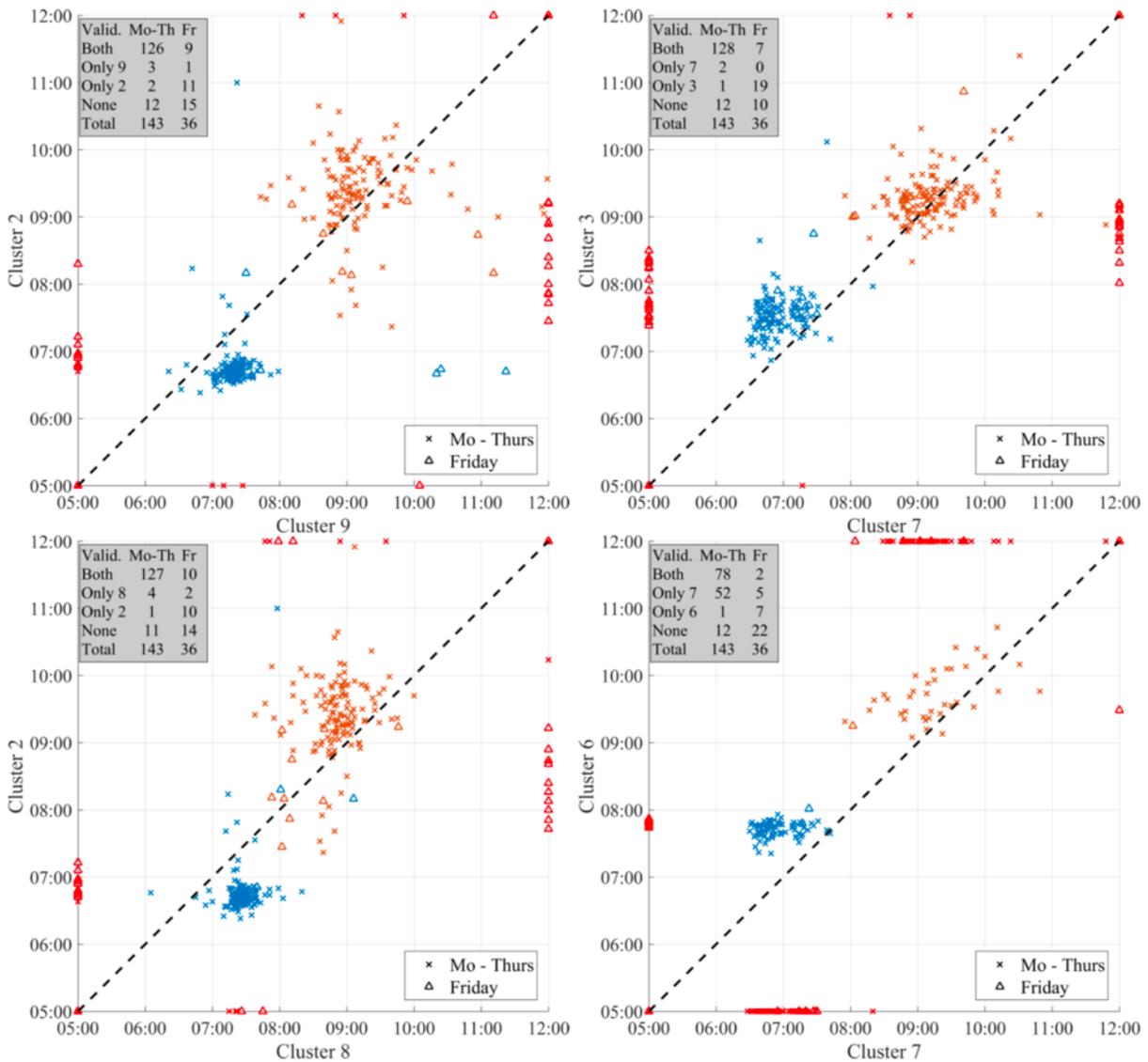

**FIGURE 8  Congestion starts and ends of several cluster combinations with significant lag. Blue markers indicate the start times and orange markers the end times of congestion. Missing congestion are visualized as symbols on the edges of the diagram.**

The most interesting cases are the ones where there is a significant lag between congestion starts and/or ends. Figure 8 illustrates four examples of pairs that show these characteristics. Observations are:

1. Cluster 2 gets congested relatively reliably between 6.30 and 7am in the morning. Cluster 8 and 9 do so at around 7.30am. Both distributions are narrow with only few outliers. This allows to formulate a statistically strong prediction rule: If cluster 2 happens to be congested between 6.30am and 7am, it is very likely that cluster 8 and 9 will get congested half an hour later.

2. In most cases cluster 7 gets congested earlier than cluster 3 and 6. Thus, also the observation of congestion in cluster 7 allows to deduce likely states for the other two clusters. However, the distribution of start times is wider and there are relatively many cases in which cluster 6 does not get congested at all.

3.  Unfortunately, the ends of congestion distribute wider than the starts, which challenges a statistical forecast. Though, in some of the presented cases there are significant lags between congestion ends as well.

Note: The depicted figures visualize only some cluster combinations. There are many more, which are left out due to limited space. Summarized, there are many pairs with less distinct distributions and some, which are totally uncorrelated (cluster 5 does not correlate with any other cluster).

Compared to the cross-correlation analysis, this type of analysis of cluster relations provides insights into daily traffic patterns of several dependent clusters. This allows to identify the distribution of starts and ends of congestion. Given narrow distributions with few outliers, this enable accurate traffic forecasts on a network level. Moreover, if lags are low, highly correlated clusters can be used to formulate expected states of the network at different times. These expectations can be fused with sparse data in order to provide more accurate current traffic estimates. Furthermore, irregularities of traffic in the network can be detected using observed traffic conditions in the clusters. However, applying these methods one has to consider that these results are based on the assumption of an invariant traffic system. If significant changes of the infrastructural supply or road usage demand occur, gathered data may become outdated. Thus, the application of such methods requires a continuous update of data. Furthermore, this analysis is performed only for the Munich road network. Similar results are expected to be found in other metropolitans, though, this proof remains for future work.

All in all, the congestion pattern analysis in this section reveals two important aspects that are relevant for traffic forecasts in networks: First, there are recurring congestion patterns that show a high degree of regularity. They can be explained well with the commuting patterns of travelers. Second, there are spatio-temporal relations between congestion at different bottlenecks of the network. For instance, some clusters are usually congested earlier than others and some clusters on alternative roads are congested at similar times.

## USING THE FEATURES FOR NETWORK-WIDE TRAFFIC STATE FORECAST
This chapter develops a traffic state predictor based on the extracted clusters. Finally, it assesses the performance of the forecast method in comparison with other methods.

### Prediction Workflow

The preceding analysis of the congestion level in the clusters revealed that there are spatio-temporal congestion patterns among the clusters that can potentially be used for improved traffic state forecast. This chapter proposes a simple data-driven method using the traffic state observed until time $t$ in order to provide a forecast for the next time steps of each cluster.

Many variables can be subject of a prediction in a traffic network. Usually, traffic speed, flow, density as fundamental variables are predicted. For travelers the most relevant information is the travel time, or, likewise, the Travel Time Loss (TTL) due to congestion. Traffic managers are interested in the reduction of TTLs of all vehicles in order to reduce the economical impacts of congestion and ensure smooth traffic flow. The instantaneous TTL on a set of edges $E^* \subseteq E$ at time $t$ is approximated as:

$$TTL_{E^*}(t) = \sum_{e \in E^*} l(e) \left( \frac{1}{V_{Rec}(e,t)} - \frac{1}{V_{Lim}(t)} \right) \tag{13}$$

Hence, one possibility is to use the TTL of an observed time interval of the current day as a feature vector for a data-driven traffic forecast. The TTL is a time series with time discretization $\Delta T$, available for each of the $n_c$ clusters. Because of a changing level of congestion and measurement inaccuracies the perceived TTL is volatile. This noise reduces the expressiveness of each individual measurement. Furthermore, for short-term predictions of a few minutes the current traffic conditions are highly relevant since congestion is relatively inert, i.e. the system requires time to change significantly. Thus, for short-term forecasts the most up-to-date measurements give important information about the traffic conditions that will prevail likely in the next minutes. Considering a longer-term forecast there is no distinct time step of the TTL which has more predictive power than any other time step. Rather, it is assumed that the overall development of the function up to the current time provides the most valuable information. Due to these reasons the TTL is aggregated over time interval $T^*$. The Summed Travel Time Loss (STTL) is defined as:

$$STTL_{E^*}^{T^*} = \sum_{t \in T^*} TTL_{E^*}(t) \tag{14}$$

The STTL represents the approximated traffic demand of edge set $E^*$ aggregated over interval $T^*$. This step further reduces the number of features while trying to keep the essential information.

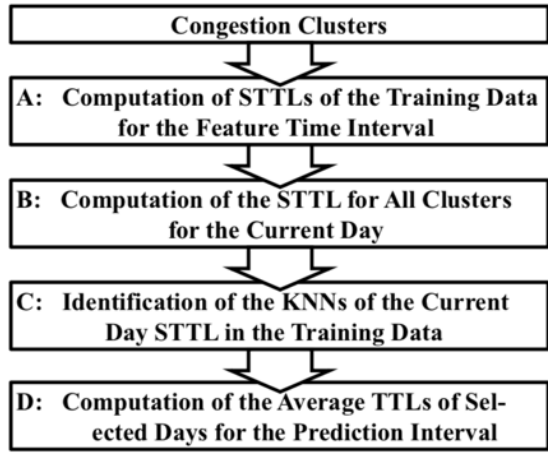

**FIGURE 9   Data-driven TTL prediction using a KNN model (see (22))**

Figure 9 summarizes the steps taken to train the model and to determine a prediction. First, the STTLs for all training days and each cluster and the specific feature time interval are computed. For the current day, the STTL of the clusters is computed for the feature time interval. Subsequently, using the current day STTLs and the ones from the training set, the $K$ most similar historical days are identified. Finally, the historical time series of the TTL of these days are averaged:

$$TTL_{d,i}^{Pred}(t) = \frac{1}{|K|} \sum_{k \in K} TTL_{k,i}(t) \tag{15}$$

Where $d$ is the index of the current day of the year, $i$ the cluster number, $K$ the set of days which have been found to be similar to the current day.

In order to select the most similar days, a simple K-Nearest Neighbors (KNN) algorithm is chosen. It involves two parameters: the number of neighbors $|K|$ that are selected for the computation of the average, and the distance metric $\Delta$, which decides about the proximity of the feature vector comprising the feature vector STTL of the current, and all historical days. In order to test the hypothesis that the TTL of a cluster can be predicted more accurately if the information of other monitored clusters is considered, the distance metric is defined as following:

$$\Delta(d_1, d_2) \coloneqq \left| \hat{R}_{X,Y}^{\gamma} \Sigma^{-1} \big( STTL_{d_1} - STTL_{d_2} \big) \right| \tag{16}$$

Where $d_1, d_2$ are two days whose STTL vector is of size $|C|$ as the number of clusters, $\Sigma^{-1}$ denotes a diagonal matrix with the reciprocal standard deviation of each of the dimensions of the STTL vector on its diagonal (used to normalize the dimensions), $\hat{R}_{X,Y}^{\gamma}$ denotes the matrix of correlations between the clusters, whose elements are raised to the power of $\gamma \in \mathbb{R}$. In this way, $\gamma$ is able to scale the influence of a correlation factor between two clusters. For instance, if $\gamma$ equals one the originally computed correlation values are used to evaluate the distance between two STTL vectors, i.e. the STTL value of two strongly correlated clusters is considered as a deciding factor. With increasing $\gamma$, all values in $\hat{R}_{X,Y}^{\gamma}$ decrease, except the ones on the main diagonal. Thus, their influence on the decision decreases. For $\gamma = 0$ all elements in the matrix turn into one, for $\gamma < 0$ it means that uncorrelated clusters gain more influence on the decision.

**Evaluation Methodology**

In order to quantify the prediction accuracy, the complete dataset is divided in a ratio of 80:20 into a training set and a test set $D_T$. The sampling is done 50 times and for each realization the following $RMSE(t)$ is computed for each forecast algorithm:

$$RMSE(t) = \sqrt{\frac{1}{|D_T||C|} \sum_{d \in D_T} \sum_{i=1}^{|C|} \big( TTL_{d,i}^{Pred}(t) - TTL_{d,i}^{GT}(t) \big)^2} \tag{17}$$

Where $TTL_{d,i}^{GT}(t)$ denominates the Ground Truth (GT) TTL at day $d$. As feature time to compute the STTLs the time from 5am to the current daytime is used for the morning traffic, for the evening clustering the time from 12.00 noon to the current daytime. For the first evaluation, the number of neighbors $|K|$ is set to 10.

**Comparison Algorithms:**

For the following comparison three variants of the KNN based predictor are considered: One, denominated as 'KNN Uni' applies a $\gamma \to \infty$, which corresponds to a correlation matrix being the identity matrix. 'KNN All' sets $\gamma = 0$ and 'KNN Cov' sets $\gamma = 10$ such that the correlation between clusters is relevant for the selection of similar historical days. As comparative algorithms three commonly applied approaches for longer-term forecasts are selected: One ('All days') takes the average of all training TTL for each time for all working days and determines the average. The second builds two clusters: One average time-series for all Mondays-Thursdays and one for all Fridays. The third further distinguishes between usual days and school holidays and thus computes four time-series in total.

**Results**

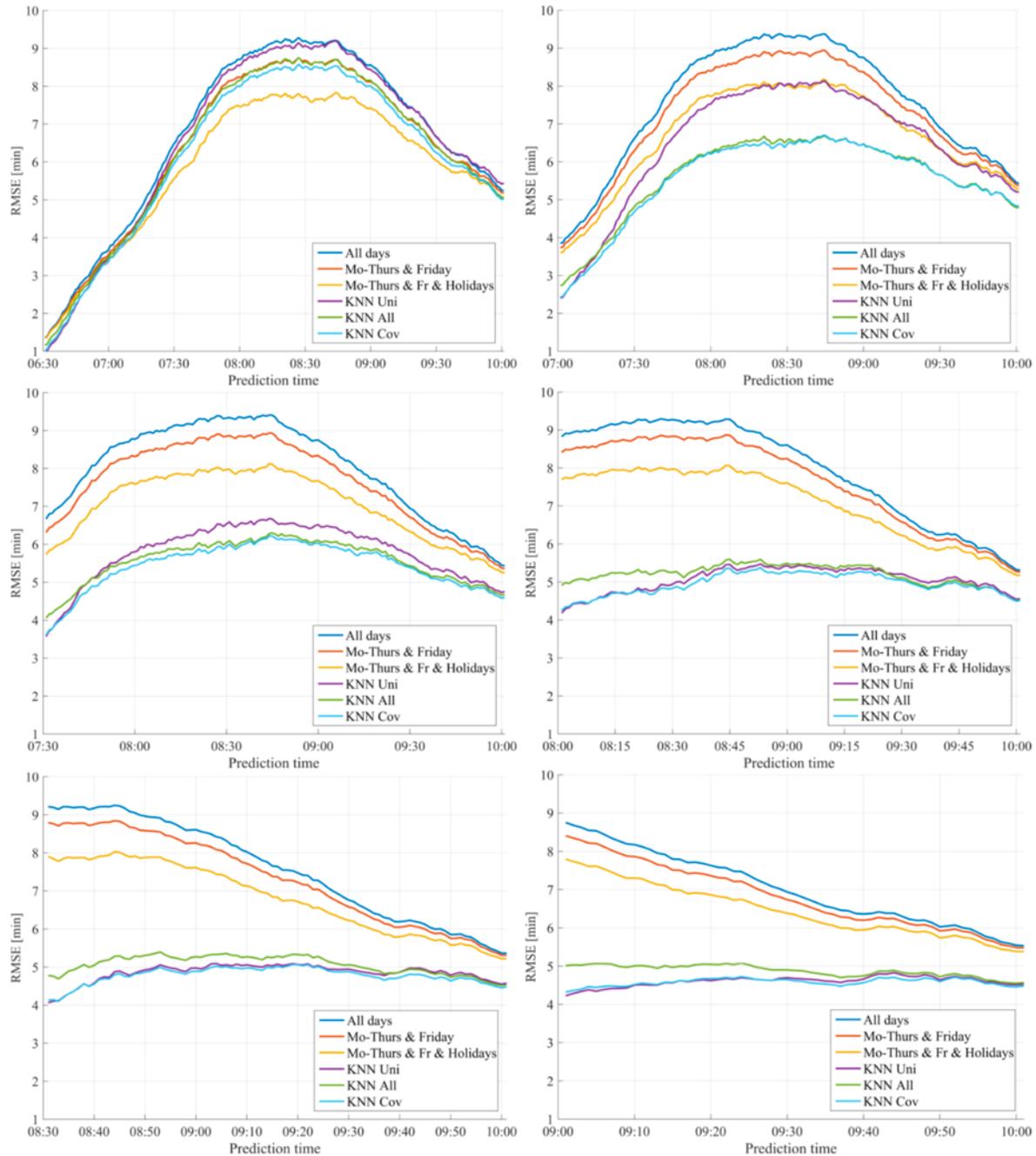

**FIGURE 10   Time-dependent forecast error of several variants of a historical average and three KNN based predictors with respect to a varying start of prediction for the morning peak**

Figure 10 depicts the RMSE for the six methods applied to different times of the day for predicting the TTLs of the morning peak until 10am. At a time of 6.30am the errors for all prediction horizons are high. The most dedicated historical average is the most accurate one. The KNN predictors are inaccurate since no congestion has been detected yet, such that no distinct prediction is possible. This changes at 7am.

While the historical average stays unaffected since they do not consider the current traffic situation, the 'KNN All' and 'KNN Cov' result in significantly more accurate predictions. Compared to the situation at 6.30am the 'KNN Uni' also improved, but less than the other KNN approaches. A similar result is illustrated in the figure depicting the errors at 7.30am. Here, the KNN predictors that consider network-wide traffic for the forecast outperform the other approaches significantly. This shows that the current network-wide level of congestion is a valuable feature for accurate traffic forecasts. At later times of prediction, the 'KNN Uni' variant improves, outperforms the 'KNN All' slightly and forecasts with similar accuracy as the 'KNN Cov'. This result can be accounted to the fact that all clusters are congested at that time such that the consideration of congestion in other clusters is decreasingly relevant. All in all, the 'KNN Cov' is the most accurate or similarly as accurate as the other KNN-based predictors. It indicates that a weighting of the feature dimensions with the cluster correlations is a way to increase the expressiveness of the features. The errors for the evening period are similar to the ones described for the morning period such that a description is omitted here.

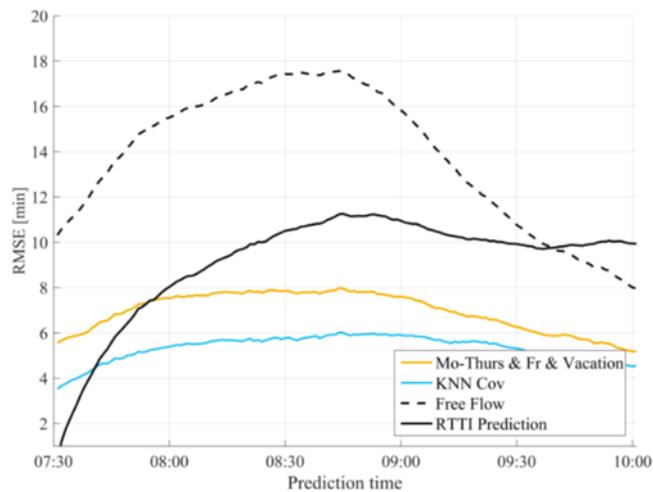

**FIGURE 11   Prediction error of a Free Flow predictor and an RTTI predictor compared to the best historical and best KNN predictor**

Figure 11 depicts the prediction error for the best historical average and the best KNN predictor in comparison to a free flow prediction (TTL = 0 for all time steps) and an RTTI prediction (the current traffic conditions are kept constant over the entire prediction period). As point in time 7.30am is chosen. As expected, the free flow prediction is the most inaccurate one with errors that exceed the other approaches over the entire prediction time. The RTTI prediction is more accurate than the other ones for a short horizon of 10-15 minutes. After this time the historical average and especially the KNN predictor are more accurate. This shows that historical averages and KNN predictors are specialized for longer term prediction horizons. Since the historical average does not consider current traffic conditions at all, and the KNN-based approaches do so only indirectly, they are not able to account for accurate short-term traffic evolutions. One simple way to gain accuracy in short-term predictions would be to fuse the current traffic situation with the longer-term prediction. The weight between these two methods could be set in such a way that it favors the current traffic conditions for a short-term horizon and favor the KNN predictor for long-term prediction horizons.

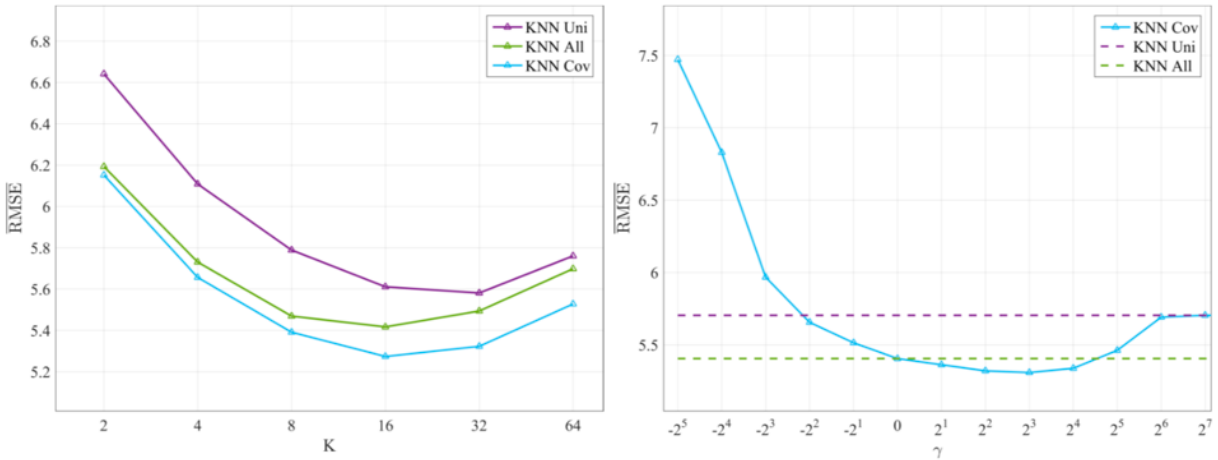

**FIGURE 12   Mean RMSE of the prediction accuracy using variants of the KNN predictor with respect to the considered number of neighbors (left) and the parameter γ (right)**

Figure 12 (left) illustrates the mean RMSE of the 'KNN Cov' predictor at 7.30am with respect to the number of considered neighbors K. The convex error function concurs with the hypothesis that too many neighbors as well as too few neighbors decrease the overall accuracy. In this case an optimal number of K ≈ 16 neighbors is found. Though, the exact number may vary depending on the time of prediction and the number of available training days. Figure 12 (right) depicts the influence of the parameter γ on the mean prediction error. Apparently there exists an optimal value for parameter γ which enables a higher prediction accuracy than the 'KNN All' and the 'KNN Uni' approach.

To conclude, if the features and coefficients are selected cautiously, the consideration of network-wide congestion for the prediction of TTLs in one congestion cluster yields the most accurate predictions. Even if γ is not optimized, the equal weighting of the features still produces more accurate results than the consideration of only one STTL value. Especially during the hours where there is substantial lag between the starts of congestion in several clusters, the presented approach outperforms other methods.

## CONCLUSION AND DISCUSSION

In this paper the question is studied, whether the network-wide traffic state is a valuable feature for a data-driven traffic state forecast method. Therefore, a network clustering approach is motivated and applied to the FCD collected from of a large fleet of vehicles over one year. The resulting static clusters are examined for congestion patterns. A cross-correlation analysis reveals strong correlations between the congestion state in different clusters and, additionally, points out temporal lags between the patterns. A subsequent study of the start and end times of congestion in clusters confirm that, with high statistical certainty, some clusters get congested earlier in a systematic way. These findings are used to formulate a simple KNN predictor for travel time losses. In a comparison of different methods the KNN approach with implemented cross-correlation outperforms other methods. This confirms that the consideration of the network-wide traffic state allows to enhance a prediction and servers as a valuable feature for a forecast method.

For future work, it is interesting to elaborate the clustering approach and study network-wide prediction in case of non-recurrent events. Furthermore, the methodology should be applied to different cities world-wide in order to confirm its generality.